\documentclass[11pt,a4paper]{article}
\usepackage{siunitx}
\usepackage[utf8]{inputenc}
\usepackage[T1]{fontenc}
\usepackage{lmodern}
\usepackage{amsmath,amssymb}
\usepackage{graphicx}
\usepackage{hyperref}
\usepackage{booktabs}
\usepackage{caption}
\usepackage{subcaption}
\usepackage{geometry}
\usepackage{authblk}

\usepackage[numbers]{natbib}
\begin{document}

\title{HotLoop Optimization of Petawatt Laser Focal Spot via a Twin-Focus Scheme}

\author{Qingfan Wu$^{1}$, Ying Gao$^{1}$, Minjian Wu$^{1}$, Jiarui Zhao$^{1,a}$, Shiyou Chen$^{1}$, 
Tianhao Liang$^{1}$, Haoran Chen$^{1}$, Tan Song$^{1}$, Zhongshuai Zhang$^{1}$, Zhangyi Wu$^{1}$, 
Shirui Xu$^{1}$, Ziyang Peng$^{1}$, Tianqi Xu$^{1}$, Zhuo Pan$^{1}$, Yujia Zhang$^{1}$, 
Qihang Han$^{1}$, Ke Chen$^{1}$, Chenghao Hua$^{1}$, Pengcheng Fan$^{1}$, Yuntian Xie$^{1}$, 
Yifei Shen$^{1}$, Shengxuan Xu$^{1}$, Liyong Ma$^{1}$, Yixing Geng$^{1}$, Chen Lin$^{1,2,3}$, 
Yanying Zhao$^{1,2,3}$, Xueqing Yan$^{1,2,3,4}$ and Wenjun Ma$^{1,2,3,4,a}$}

\affil{$^{1}$State Key Laboratory of Nuclear Physics and Technology, School of Physics, Peking University, Beijing 100871, China}
\affil{$^{2}$Beijing Laser Acceleration Innovation Center, Huairou, Beijing 101400, China}
\affil{$^{3}$Institute of Guangdong Laser Plasma Technology, Baiyun, Guangzhou 510540, China}
\affil{$^{4}$Collaborative Innovation Center of Extreme Optics, Shanxi University, Taiyuan, Shanxi 030006, China}

\affil{$^{a}$) Authors to whom correspondence should be addressed: wenjun.ma@pku.edu.cn, jrzhao@pku.edu.cn}

\date{}

\maketitle

\begin{abstract}
Achieving diffraction-limited focusing of high-power laser pulses to generate ultra-high intensities is crucial for developing compact laser-driven particle accelerators and exploring strong-field quantum electrodynamics. However, accurately diagnosing and optimizing the focal spots of petawatt (PW) laser pulses remains a significant challenge. In this work, we present an experimental methodology utilizing a twin-focus scheme to precisely characterize the intensity distribution and wavefront of focused PW femtosecond laser pulses, and employ it to elucidate their power-dependent evolution. Furthermore, we optimized the focal spots at full power via our in-situ wavefront correction method termed ``HotLoop'', achieving a Strehl ratio of 0.80 for 1 PW laser pulses. Consequently, the cutoff proton energies in laser proton acceleration experiments were significantly enhanced. The success of this approach underscores the necessity of in-situ high-energy wavefront correction for ultra-high intensity laser-matter interactions.
\end{abstract}


\section{Introduction}

Ultra-high intensity lasers have undergone rapid development over the past two decades \cite{1,2}. Serving as powerful tools, they have enabled the investigation of extreme phenomena in strong-field quantum electrodynamics\cite{3,4,5}, and facilitated relativistic laser-matter interaction experiments, such as electron acceleration\cite{6}, ion acceleration\cite{7,8,9}, and X/$\gamma$-ray generation\cite{10,11}. In these frontiers, the achievable peak intensity at the focus acts as the governing metric\cite{12,13,14}. For instance, in laser-driven proton acceleration, proton cutoff energies scale critically with peak intensity. An on-target intensity exceeding $10^{21}$ W/cm$^2$ is essential to generate protons over 100 MeV for tumor therapy\cite{15,16,17}. To reach such extreme intensity, adaptive optics (AO) systems employing deformable mirrors (DM) are routinely used to correct laser wavefront aberrations, enabling diffraction-limited focusing. Typically, the AO wavefront correction protocols utilize low-power seed beams for closed-loop optimization to eliminate static aberrations\cite{18,19}, failing to compensate for dynamic distortion during high-power operations. Despite recent advances in high-power wavefront diagnostics and correction at several facilities\cite{13,14,20,21}, these investigations are seldom integrated with the actual laser-target interaction. In practice, the critical alignment between the laser and the target still relies on low-power beams, which ultimately limits the practical realization of ultra-high intensities in experiments.

There are multiple sources of aberrations that would degrade the laser focal spots in high-power laser systems. During low-power operation, static aberrations arising from optical imperfections are the dominant factors reducing peak intensity of the focal spots; these are nevertheless correctable via deformable mirrors at low power. In contrast, when the laser operates at high power, additional sources of aberrations emerge, leading to a significant discrepancy between the optimized low-power wavefront and the actual high power operational state. The primary contributors include thermal lensing induced by pump heat loads in the gain medium\cite{22,23}, and in high-repetition-rate systems, thermal deformation of compressor gratings \cite{24,25}. Additionally, wavefront modulation induced by plasma mirrors (PM), which is employed for improving temporal contrast\cite{26,27}, acts as a further source of aberrations. Unlike static aberrations, these aberrations are immeasurable during low-power operation and thus cannot be corrected accordingly.

Therefore, characterizing and optimizing the focal spots of the full power laser beam is essential to achieve the highest possible intensity\cite{18}. A common approach involves extracting the leaked pulses from the rear surface of a mirror through a vacuum window, then relays them with focusing optics for characterization\cite{19}. This method is highly susceptible to distortions from transmissive optics and atmospheric turbulence. Measurements performed in vacuum have shown superior results. However, the challenges in implementation are significant. Alternatively, Yoon et al. used wedge-shaped quartz plates placed upstream of the compressor to attenuate the beam for thermal lens characterization\cite{18}, but it was unable to measure aberrations introduced by the PMs. Park et al. reported that random pinhole attenuators (a board featuring numerous randomly distributed pinholes) can effectively attenuate beam energy and preserve the low-spatial-frequency information of the wavefront. This method has been used to measure 1 PW focal spots on the CoReLS laser system, albeit at the expense of high-spatial-frequency resolution\cite{28}.

Besides direct measurements, several indirect strategies for assessing focal spots' intensities have been proposed. For instance, Vishnyakov et al. utilized third-harmonic reflection and on-axis hard X-ray diagnostics to identify the focal position of the maximum intensity\cite{29}. Xu et al. recently showed that the far-field pattern of the second harmonic emitted during laser-foil interactions correlates with the shape of the ultra-intense laser focus and can thus serve as an indirect diagnostic tool for high-energy focal spots\cite{30}. Although these indirect methods allow one to determine the maximum peak intensity, they do not provide direct access to wavefront information and thus cannot be used for active wavefront correction.

To mitigate wavefront aberrations at high power, a closed-loop strategy that operates under actual operational conditions is essential. Conceptually, it should integrate an in situ, non-invasive, real-time wavefront characterization with immediate feedback to the deformable mirror. We refer to this high-power-compatible focus-optimization protocol as ``HotLoop''. However, the experimental realization of this approach has not yet been reported owing to the inherent difficulties in characterizing the focal spots at ultra-high intensities.

In this work, we developed a twin-focus adaptive optics (TFAO) system specifically designed to characterize and optimize the focal spots of PW femtosecond laser pulses. Utilizing this system, we generated a high-fidelity, highly attenuated replica (the ``twin focus'') of the full-energy focal spot. By characterizing the twin focus, the properties of the full-energy focal spots were learned. We found that the high-power wavefront distortions of the focal spots were significantly more severe than those observed at low power, necessitating the use of HotLoop for optimal performance. To guide aberration correction, we conducted a detailed investigation into the power-dependence of thermal aberrations, comparing measurements both with and without a plasma mirror. By integrating the TFAO system into our closed-loop strategy, we successfully implemented HotLoop optimization on 1 PW laser pulses and produced a diffraction-limited focal spot with a Strehl ratio (SR) of 0.80. Finally, the proton acceleration results indicated that the optimized focal spots led to a significant enhancement in proton energy, directly validating the efficacy of our scheme.

\section{Experiment Setup}
The experiments were performed on a 1 Hz/30 fs 2 PW Ti:sapphire laser system hosted at the CLAPA-II facility\cite{1,31,32}. It consists of two Front-End amplifiers (XPW (100Hz) + OPCPA (10Hz)), a grating pulse stretcher, five multi-pass amplifiers, and a PW grating compressor. Before the compressor, the beam is expanded to a diameter of 330mm to accommodate a maximum design output of 60J (see Fig. 1).

\begin{figure}[htbp]
\centering
\includegraphics[width=0.9\textwidth]{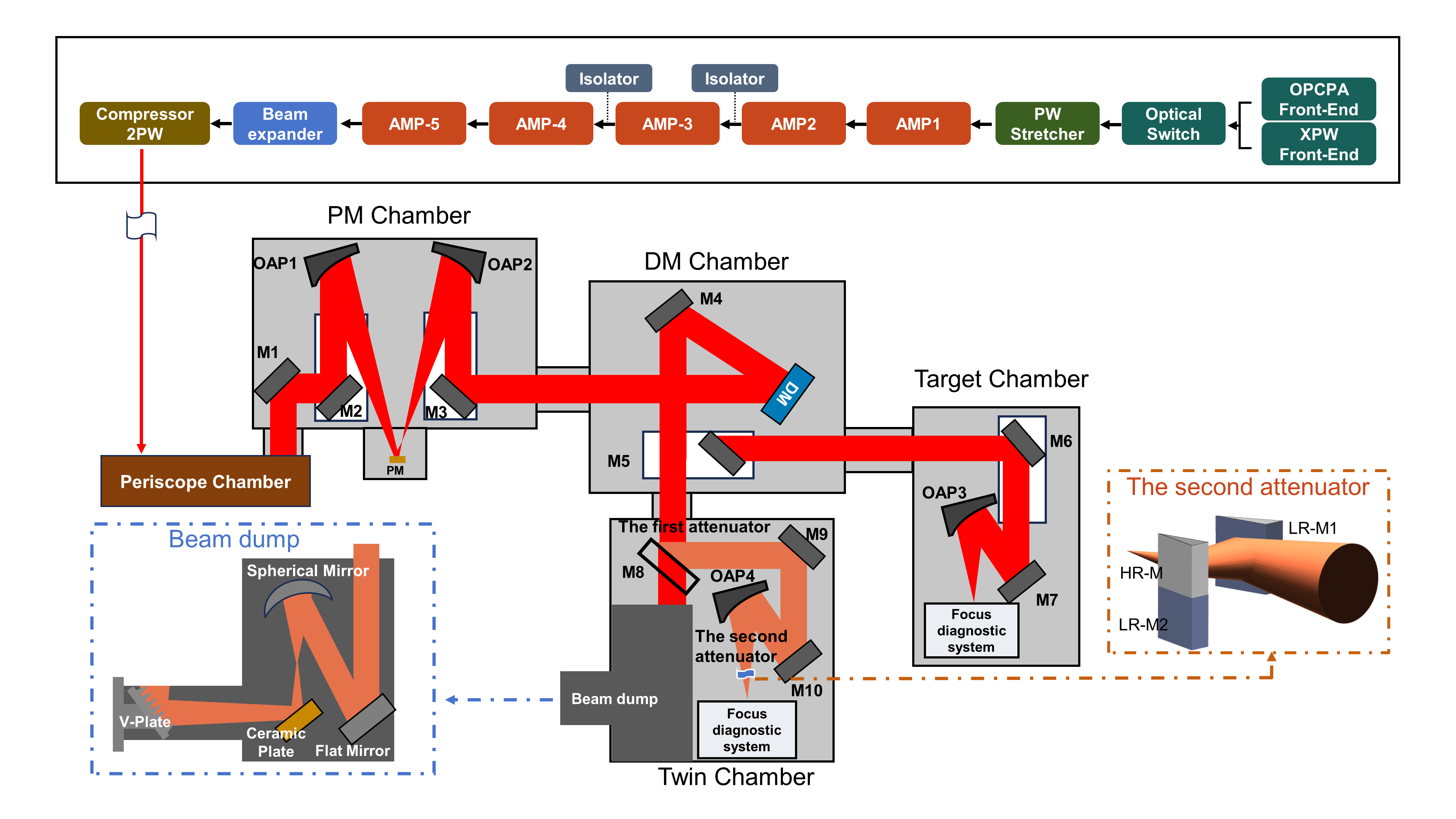}
\caption{Layout of the PW laser and the experimental setup. }
\label{fig:laserbeam}
\end{figure}

In the experimental area, the compressed laser beam first enters a Periscope Chamber for polarization conversion, resulting in an S-polarized output. It then proceeds to a PM system for temporal contrast enhancement at a trigger fluence of $200\ \text{J/cm}^2$ and a reflectivity of 80\%.
Subsequently, it passes a 500-mm diameter mechanical deformable mirror (MD series, ISP System, France) with 129 precision actuators for the compensation of the aberrations both intrinsic to the laser and introduced by the beamline. Finally, the pulse is delivered into the Target Chamber and focused by a f/3 parabola mirror for laser-matter interaction.

In the Target Chamber, the focal spots are characterized by a custom-built vacuum-compatible focus diagnostic system (Fig. 2(a)(b)). It contains a near-field camera and two far-field cameras (GED200M, Mindvision) for beam alignment and focus intensity measurements, along with a wavefront sensor (Sid4-V, Phaics) positioned at the optical relay plane conjugate to the DM. To prolong the sensor's operational time in vacuum, we utilize copper heat sinks attached to the chips of the CCD for cooling. The wavefront sensor can be used for closed-loop wavefront correction in the low-power mode ($1\ \text{mJ}$). After the correction, the full width at half maximum (FWHM) and the Strehl ratio (SR) of the optimized focal spots were $3.5 \times 3.2\ \mu\text{m}$ and 0.9, respectively, indicating a near-diffraction-limited performance. The SR was derived from the wavefront RMS measured by the wavefront sensor.

\begin{figure}[htbp]
\centering
\includegraphics[width=0.95\textwidth]{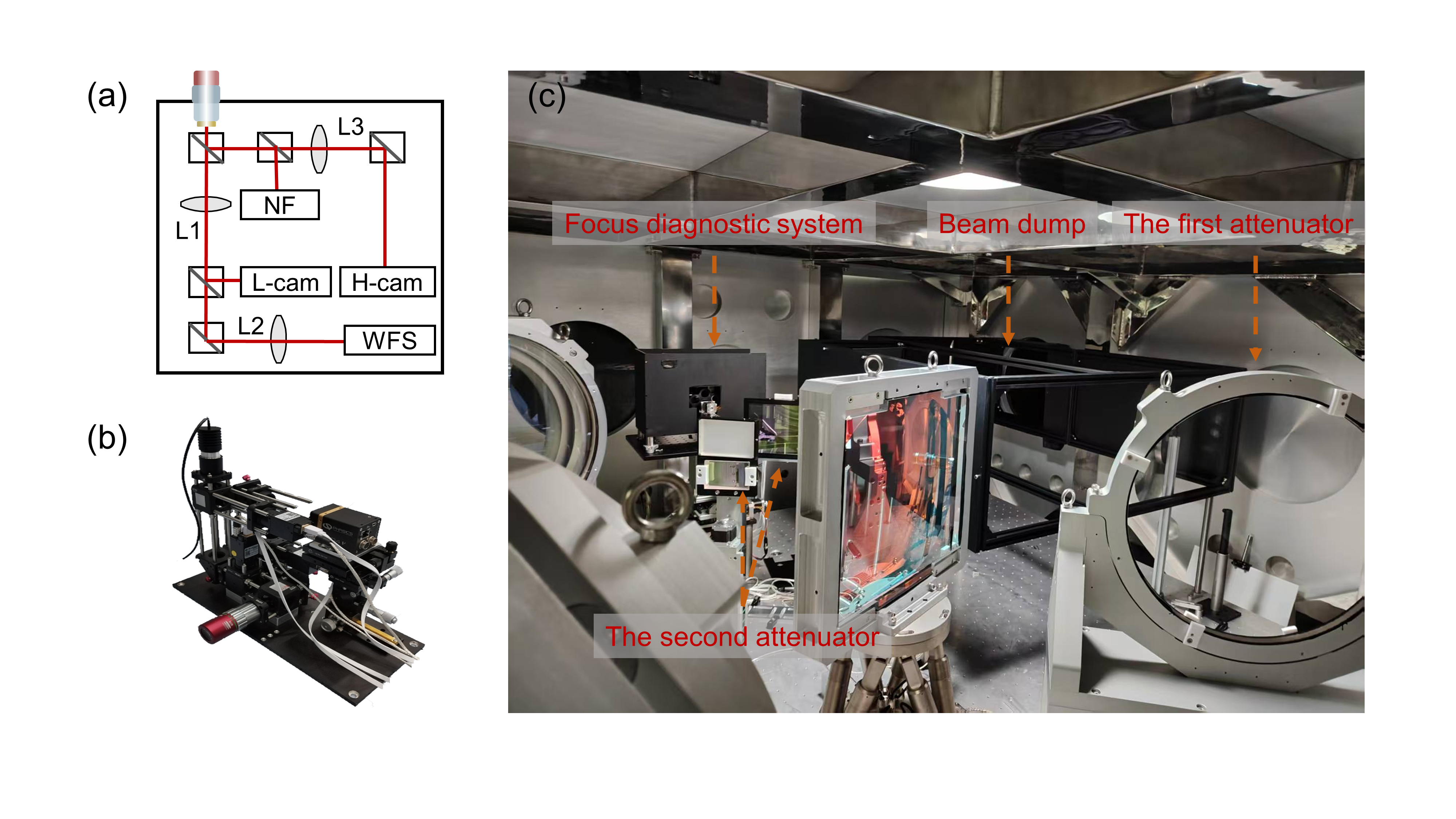}  
\caption{(a) Optical layout of the focus diagnostic system; (b) Focus diagnostic system; (c) Overview photo of the Twin Chamber.}
\label{fig:focus_diagnostic}
\end{figure}

To perform focal spot characterization and HotLoop optimization in the high-power mode (laser energy up to 30 J), we constructed the TFAO system incorporating with a ``Twin Chamber'' (see Fig. 2(c)) coupled to the Target Chamber (see Fig. 1). The Twin Chamber houses a focusing off-axis parabolic (OAP) mirror identical to the one in the Target Chamber (i.e., focal length of 990 mm and off-axis angle of $45^\circ$). The twin OAPs and the large mirrors were fabricated in the same batch, which is essential for the implementation of the Twin-focus scheme, their surface RMS errors are $0.019\ \lambda$ and $0.016\ \lambda$, respectively. A two-stage all-reflective attenuation system was utilized to attenuate pulse energy by 5 orders of magnitude for characterization. The first attenuator employs a 500 mm uncoated fused silica mirror (M8), which transmits approximately 84\% of the energy into a beam-dump (see inset of Fig. 1). Within the beam dump, a ceramic plate diffusely reflects the incident energy onto the surrounding metal piping, where it is passively cooled via natural convection. This configuration is specifically engineered for sustained 1-Hz operation at 60 J. The residual beam is further attenuated by the second attenuator consisting of a fixed 0.1\% reflectivity mirror (LR-M1) and a motorized dual-zone mirror (see inset of Fig. 1). The latter is composed of a high-reflectivity section (HR-M, 98\%) and a low-reflectivity section (LR-M2, 0.1\%), selectable for focus characterization in the low- and high-power modes, respectively. Since all attenuators are reflective, nonlinear phase accumulation (B-integral) is avoided, thereby preserving wavefront fidelity. This system provides the technical basis for producing a high-similarity replica of the high energy spot in the Target Chamber, which is essential for enabling the implementation of characterization and optimization of the focal spots at high power.

\section{Methods}

\subsection{Generation of the twin focus}

The characterization of the high-power focal spot relies on establishing a highly attenuated and high-fidelity replica of the Target Chamber's focal spot inside the Twin Chamber. To achieve this, we first generated a near-diffraction-limited focal spot (the primary focus) in the Target Chamber via closed-loop wavefront correction in the low-energy mode. Subsequently, we generated a comparable focal spot (see Fig.~3(a)(g)) in the Twin Chamber by aligning OAP4 with the same DM settings. Since all the optical components in the two chambers are identical, the low-energy replica in the Twin Chamber closely matches the high-power spot in both shape and intensity distribution. This correspondence ensures that the twin focus and the primary focus respond consistently to specific wavefront aberrations. To validate this, we deliberately introduced specific Zernike aberrations via the DM. The resulting deformed focal spot in both chambers are depicted in Fig.~3, corresponding to $0.5\lambda$ astigmatism ((b)(h) and (c)(i)), $0.5\lambda$ coma ((d)(j) and (e)(k)), and $0.5\lambda$ trefoil ((f)(l)). The fidelity between the two focus images is evaluated using the Structural Similarity Index (SSIM) and Mean Squared Error (MSE), defined as
\begin{equation}
    \text{SSIM}(x,y) = \frac{(2\mu_x\mu_y + c_1)(2\sigma_{xy} + c_2)}{(\mu_x^2 + \mu_y^2 + c_1)(\sigma_x^2 + \sigma_y^2 + c_2)},
    \label{eq:ssim}
\end{equation}
\begin{equation}
    \text{MSE} = \frac{1}{mn} \sum_{i=1}^{m} \sum_{j=1}^{n} \left[ I_x(i,j) - I_y(i,j) \right]^2,
    \label{eq:mse}
\end{equation}

where the subscripts $x$, $y$ denote the two images to be compared in the structural similarity calculation, $\mu_x = \frac{1}{N} \sum_{i=1}^{N} x_i$ and $\mu_y = \frac{1}{N} \sum_{i=1}^{N} y_i$ represent the mean intensity of the two compared images, $\sigma_x = \sqrt{\frac{1}{N} \sum_{i=1}^{N} (x_i - \mu_x)^2}$ and $\sigma_y = \sqrt{\frac{1}{N} \sum_{i=1}^{N} (y_i - \mu_y)^2}$ denote the corresponding standard deviations, and $\sigma_{xy} = \frac{1}{N} \sum_{i=1}^{N} (x_i - \mu_x)(y_i - \mu_y)$ represents the covariance between the two images. Additionally, the constants $C_1 = (K_1 L)^2$ and $C_2 = (K_2 L)^2$ are introduced as small stability regularization terms to avoid division-by-zero numerical singularity in extremely uniform or low-intensity regions, with standard settings $K_1 = 0.01$, $K_2 = 0.03$, and $L = 255$ for 8-bit images in this work. In equation (2), $I_x(i,j)$ and $I_y(i,j)$ correspond to the pixel values at a given position in the two images.

SSIM comprehensively quantifies image similarity across three perceptual dimensions: luminance, contrast, and structural correlation. As a similarity-focused metric, SSIM ranges from $-1$ to $1$, with values approaching $1$ indicating excellent consistency in visual features and structural details between paired images. In contrast, MSE calculates the average squared pixel-wise intensity difference between two images. As an error-based metric, a smaller MSE indicates a lower pixel discrepancy. Fig. 3 presents a comparison of focal spots between the Twin Chamber (a--f) and Target Chamber (g--l) under identical prescribed aberrations. It is evident that the focal spots in both chambers show excellent consistency. Table 1 reveals that the SSIM values for all aberration modes exceed $0.92$, confirming the high fidelity of the replica.

\begin{figure}[htbp]
\centering
\includegraphics[width=0.95\textwidth]{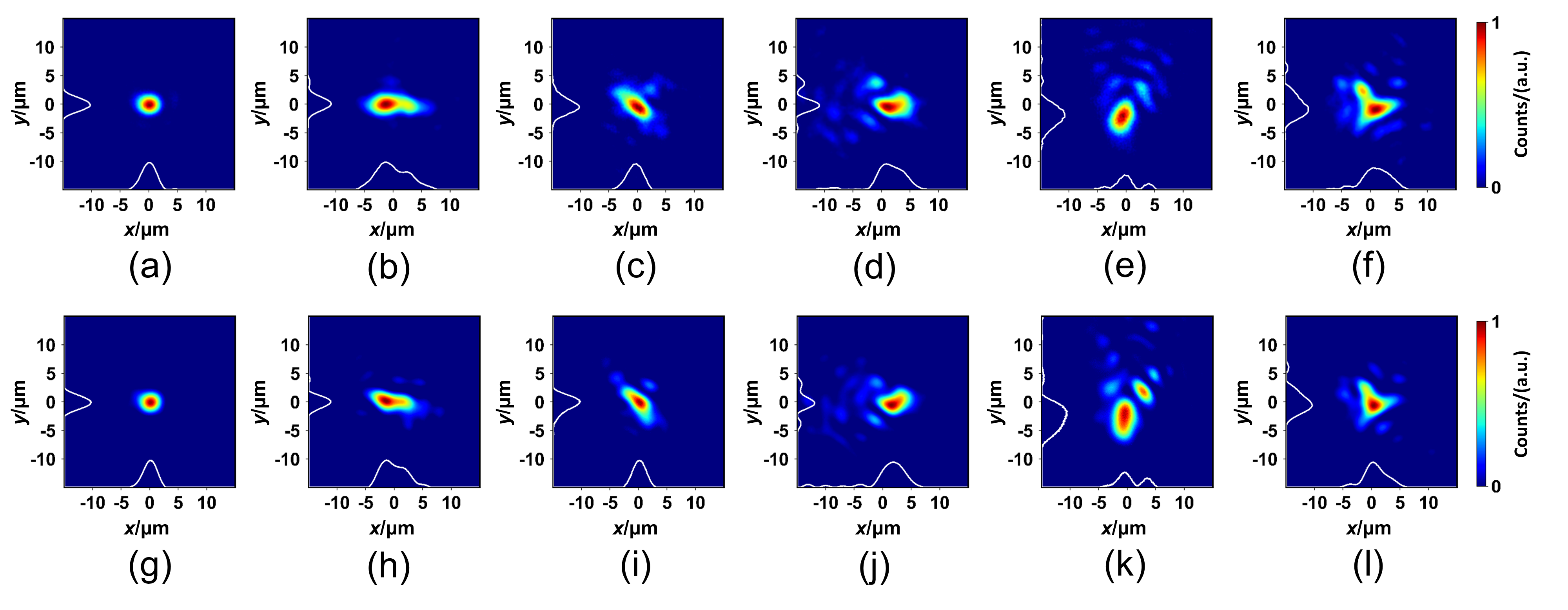}
\caption{Comparison of focal spots between the Twin Chamber (a-f) and Target Chamber (g-l) under identical prescribed aberrations.(The colorscale is the same for all frames): (a, g) flat wavefront, (b, h) $0^\circ$ astigmatism, (c, i) $45^\circ$ astigmatism, (d, j) Coma X, (e, k) Coma Y, (f, l) $0^\circ$ trefoil.}
\label{fig:focal_comparison}
\end{figure}

\begin{table}[htbp]
\centering
\caption{Customized wavefront and similarity of the twin focus and the main focus}
\begin{tabular}{ccccccc}
\hline
Parameter & Ideal Flat & \shortstack{Astigmatism \\ $0^\circ$} & \shortstack{Astigmatism \\ $45^\circ$} & Coma X & Coma Y & \shortstack{Trefoil \\ $0^\circ$} \\
\hline
Zernike coefficients & / & +0.5 & +0.5 & +0.5 & +0.5 & +0.5 \\
SSIM & 0.943 & 0.968 & 0.924 & 0.947 & 0.933 & 0.933 \\
MSE & 0.00007 & 0.002 & 0.007 & 0.0017 & 0.008 & 0.0009 \\
\hline
\end{tabular}
\label{tab:wavefront_similarity}
\end{table}

\subsection{HotLoop Focus Optimization}

Based on the consistency between the twin focus and the main focus, we utilize the wavefront measured from the attenuated twin focus to drive the DM, thereby simultaneously optimizing the high-power focus. The standard low-power wavefront correction protocol comprises three stages: establishing optical conjugation between the WFS and the DM, calibrating the DM response, and executing the closed-loop optimization based on real-time WFS wavefront measurements. The first two stages of the HotLoop optimization protocol are identical to those of the standard routine. Instead, after the DM calibration, the attenuation selector in the Twin Chamber is switched from low- to high-energy mode. Crucially, this transition must introduce negligible displacement or deformation to the focal spots. To achieve this, the co-alignment was initially optimized using a 300 mm aperture expanded pilot beam from an 808 nm laser diode, followed by a final in-situ verification using the low-energy main beam. Finally, real-time closed-loop optimization is performed at full power using the feedback from the attenuated replica. It should be noted that the success of the HotLoop relies on the validity of the low-power calibration file during high-power states. This imposes stringent requirements on both the mechanical repeatability of the attenuation selector switching and the fidelity of the twin-focus configuration.

\subsection{Inclusion of the Plasma Mirror}

In many PW laser experiments such as laser ion acceleration\cite{33,34,35}, it is necessary to include PMs to enhance the temporal contrast. They are triggered only when the fluence exceeds $\sim 10\ \text{J/cm}^2$, thereby filtering the low-energy pre-pulses while efficiently reflecting the main pulse. It has been found that the PMs would introduce additional wavefront aberrations because of the non-uniform modulation of the plasma's critical-density surface. Therefore, it is essential to include the PMs in HotLoop protocol if they are employed. To date, robust closed-loop optimization for 1 PW focal spots downstream of a PM has not been reported.

For accurate characterization and optimization of the focal spots downstream of the PM, the workflow must be specifically adjusted. Since the PM surface is refreshed (translated to a new position) after each shot, it is essential to decouple the aberrations resulting from the mechanical errors and that from those caused by plasma modulation. Therefore, we implement a pre-verification protocol: before full-power operation, we verify that the focal spot's position and profile remain invariant during the translation of the PM. This ensures that any wavefront distortions measured during high-power operation are strictly attributable to the plasma triggering process rather than mechanical instabilities.

\section{Experimental Results}

\subsection{Focal Spots and Wavefront of 1PW laser pulses with PM}

\begin{figure}[htbp]
\centering
\includegraphics[width=0.95\textwidth]{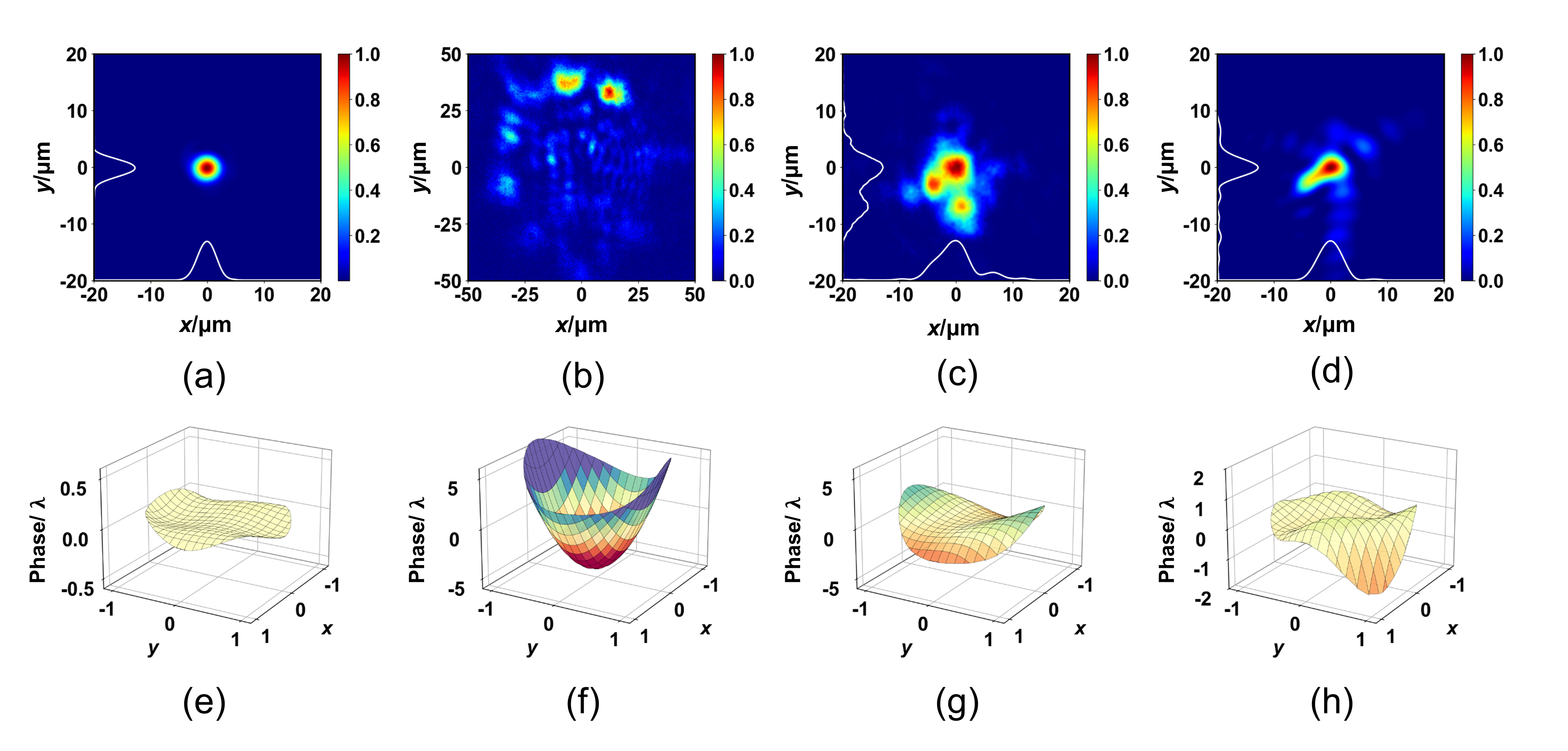}
\caption{Comparison of focal spots and wavefronts at different energy levels and compensation stages in the Twin Chamber. (a, e) Measured at 1 mJ (base loop); (b, f) Measured at 30 J without compensation; (c, g) Measured at 30 J after compensating for the defocus term; (d, h) Calculated spot and measured wavefront for 30 J, obtained after compensating for both defocus and astigmatism terms.}
\label{fig:focus_wavefront_pm}
\end{figure}

To illustrate the impact of thermal aberrations on the laser focal spot with PM, the 1PW laser's focal spot and wavefront were measured in the Twin Chamber using the TFAO system. Fig. 4(a) and Fig. 4(e) present the baseline performance at 1 mJ, where AO correction yielded a near-diffraction-limited spot with a SR of 0.9 (calculated from the wavefront RMS). In contrast, the 1 PW regime exhibits remarkable degradation in both intensity distribution (Fig. 4(b)) and wavefront quality (Fig. 4(f)). Specifically, we observed a forward focal shift accompanied by the enlargement of the focal spot, primarily driven by a dominant defocus term in the high-energy wavefront (Fig. 4(f)). The focal position was found to shift by 400 $\mu$m longitudinally towards the OAP mirror due to this defocus term, indicating a pronounced pre-focusing of the beam at high power. It can be compensated for by defocusing the OAP, and the compensated focal spot is shown in Fig. 4(c). While mechanical OAP adjustment can mitigate this defocus, the residual focal spot quality (SR=0.14, calculated from the wavefront RMS) remains inferior to the optimized low-power focal spot. Wavefront decomposition shows that this residual degradation stems from the amplification of other aberrations (Fig. 4(g)). Although the astigmatism can be eliminated by adjusting the OAP, such iterative optimization at full power in PW experiments is shot-consuming and potentially dangerous. Therefore, we choose to directly compensate for the wavefront distortion by using the deformable mirror. Fig. 4 (d) presents the calculated results where both defocus and astigmatism are eliminated, representing the theoretical limit of the OAP-adjusting scheme. It's clear that although the focal spot is significantly improved, the residual aberrations (e.g., trefoil, coma, and spherical aberration) in the wavefront (see Fig. 4(h)) still notably degrade the peak intensity of the focal spot.

\subsection{Power-dependent evolution of aberrations in the presence and absence of a plasma mirror}

\begin{figure}[htbp]
\centering
\includegraphics[width=0.95\textwidth]{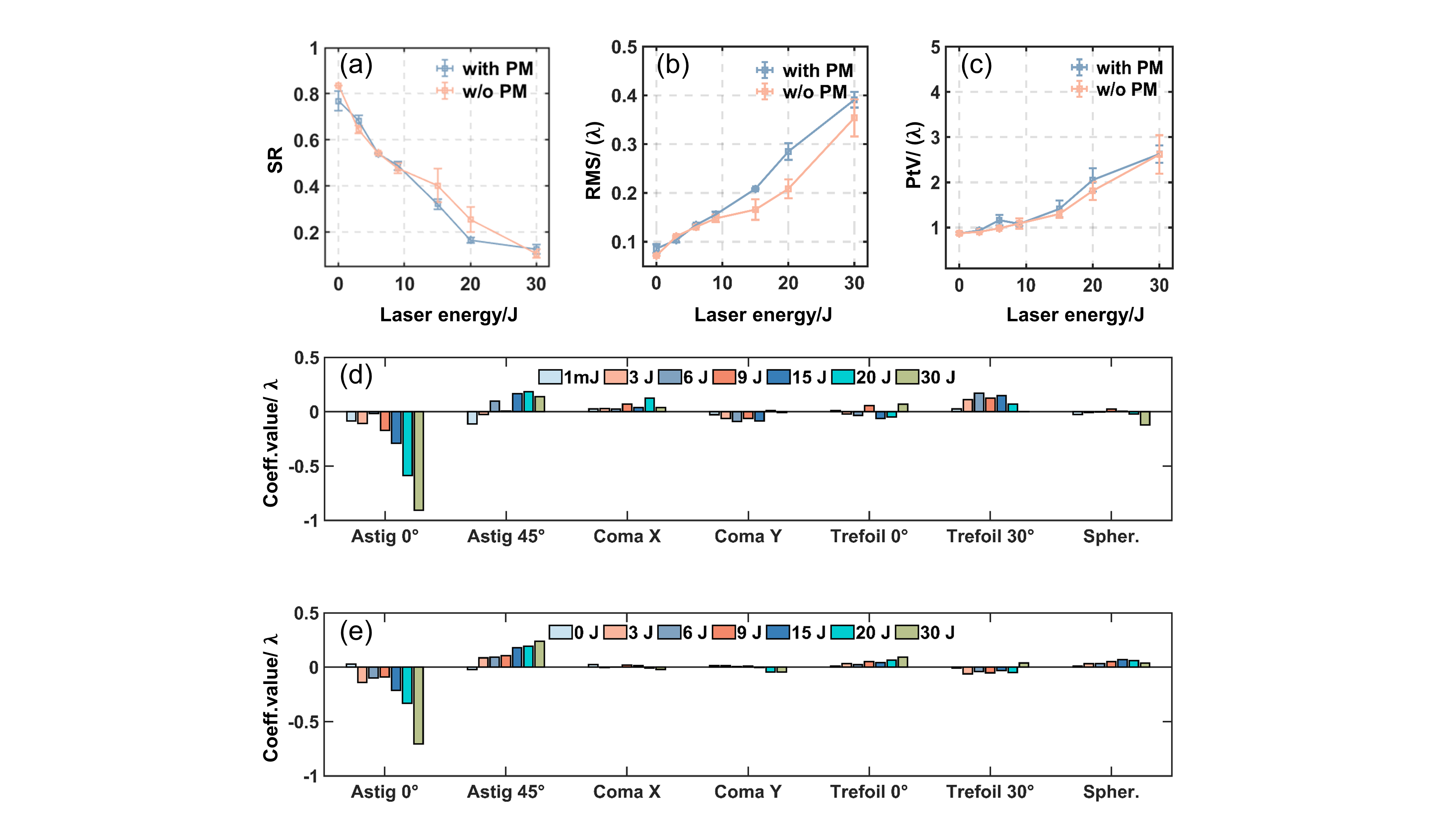}
\caption{The energy-dependent evolution of the wavefront. (a) The dependency of SR; (b) RMS of wavefront; (c) PV of wavefront; (d) Aberration coefficients with PM; (e) Aberration coefficients without PM.}
\label{fig:wavefront_evolution}
\end{figure}

To characterize the energy-dependent evolution of the focal spots, we systematically analyzed the intensity distribution and wavefront quality for laser energy ranging from 1 mJ to 30 J after the compressor, comparing configurations with and without the PM. As illustrated in Fig. 5(a), the SR exhibits a monotonic decline as the laser energy scales up. Correspondingly, the root-mean-square (RMS) wavefront error and peak-to-valley (PV) values generally increase with energy (see Fig. 5(b)(c)). While wavefront aberrations deteriorated in both configurations, the integration of the PM system led to further degradation compared to the setup without the PM.

Fig. 5(d) and Fig. 5(e) present a detailed decomposition coefficients of aberrations across various energy levels, revealing that distinct aberration modes exhibit specific sensitivities. Notably, the coefficients for the dominant aberrations, including coma, astigmatism, and trefoil, demonstrate a quasi-linear growth with increasing energy. This phenomenon can be attributed to two reasons: (1) The oblique incidence of the pump beam creates an elliptical heating profile on the crystal. This produces an anisotropic refractive index gradient, causing the thermal lens to lose its circular symmetry. (2) The oblique incidence of the seed beam introduces off-axis aberrations analogous to lens-induced distortions, which scale progressively as the energy increases. These results establish a direct correlation between laser energy and the aberrations, which holds significant implications for high-power laser systems subject to stringent wavefront requirements.

\subsection{HotLoop optimization of PW laser focal spots with PM}

After the characterization of the focal spots, HotLoop wavefront correction was performed at 30 J. Leveraging the precise initial calibration provided by the TFAO system, the HotLoop converged to its optimal state within only six shots (Fig.~6(a)). This rapid convergence is attributed to the high-fidelity synchronization between the low-energy and high-energy correction protocols.

To further quantify the HotLoop's performance, we analyzed the encircled energy profiles shown in Fig.~6(b). The results indicate that thermal aberrations significantly degrade the energy concentration at high power when the DM adopted the 1\,mJ wavefront correction setting. In contrast, the HotLoop effectively compensated for the thermal wavefront distortions, substantially enhancing the energy concentration at 30\,J. Notably, the corrected high-power focal spot (Fig.~6(c)) exhibited a FWHM of $3.38 \times 3.29\ \mathrm{\mu m}$ with 47\% energy within this area, nearly reaching the theoretical diffraction limit. By comparing the effective areas ($A_{eff}$) between the calculated ideal diffraction-limited focal spot and our experimentally measured 30 J focal spot (Fig.~6(c)), we determined the Strehl ratio to be 0.73. This value is slightly lower than the wavefront RMS-derived SR of 0.80.

\begin{figure}[htbp]
\centering
\includegraphics[width=0.95\textwidth]{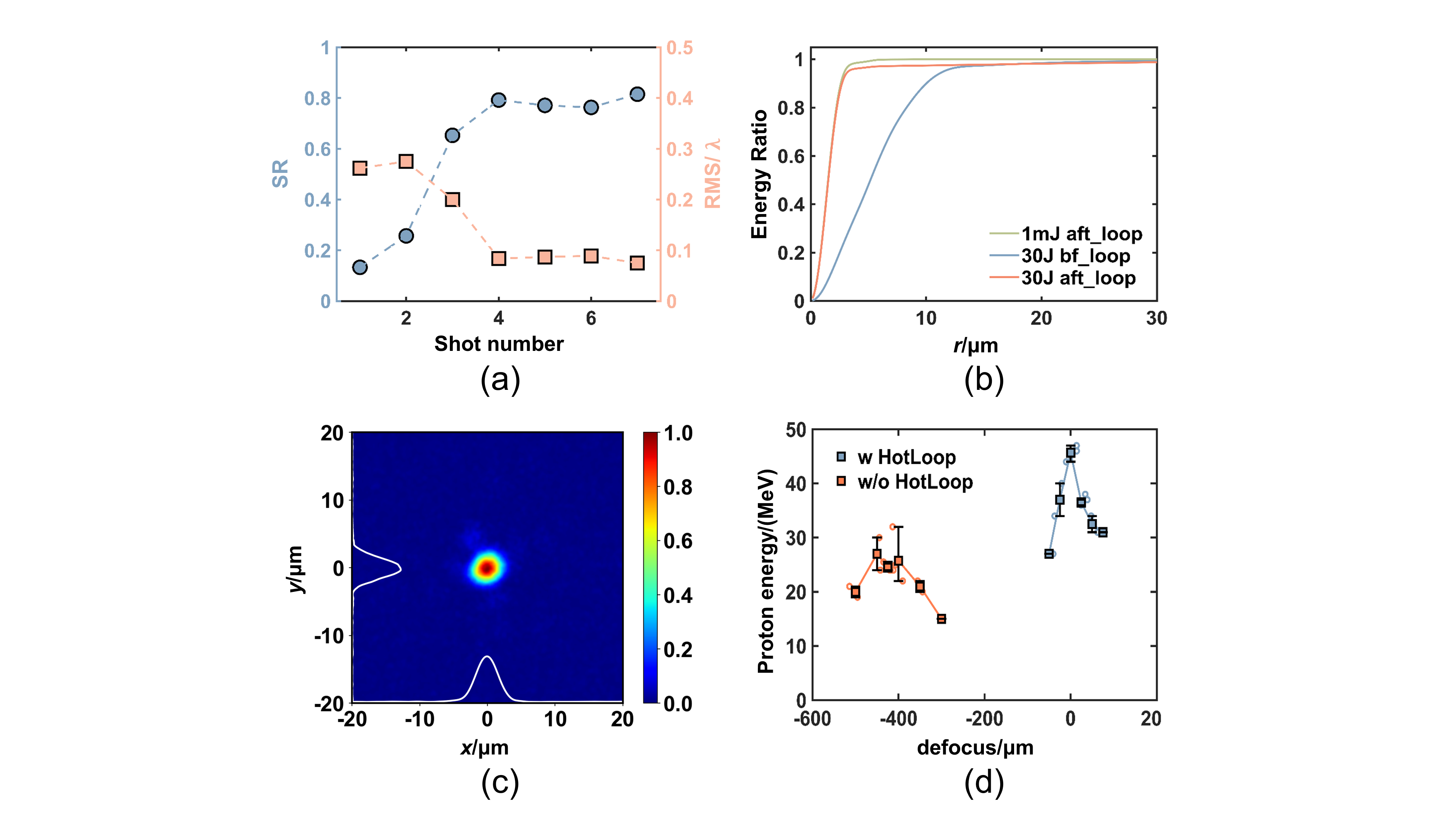}
\caption{Wavefront correction of high-energy laser focal spots. (a) Evolution of the SR during the HotLoop wavefront correction process at $30\ \text{J}$; (b) Encircled energy ratio comparison between the $1\ \text{mJ}$ and $30\ \text{J}$ loop; (c) Measured full-bandwidth focal spot at $30\ \text{J}$ after the HotLoop; (d) Comparison of proton cut-off energy of $1\ \mu\text{m}$-thick targets.}
\label{fig:hotloop_correction}
\end{figure}

To evaluate the efficacy of the HotLoop strategy, ion acceleration experiments were conducted using $1\ \mathrm{\mu m}$-thick polymer targets. Specifically, we compared the proton energy corresponding to focal spots with and without HotLoop optimization. P-polarized laser pulses (21 J, 30 fs FWHM) were focused onto the target at $10^\circ$ incident angle after passing through a single PM system. The proton cutoff energy was measured using a Thomson parabola spectrometer (TPS) in the experiment. While the peak intensity is calculated as $3.1 \times 10^{21}\ \mathrm{W/cm^2}$ based on the intensity distribution of the optimized low-power focal spot (Fig.~4(a)), thermal-induced aberrations at high power would typically degrade the focus and reduce the actual peak intensity to $1.15 \times 10^{21}\ \mathrm{W/cm^2}$ (Fig.~4(c)). By implementing the HotLoop optimization to counteract these effects, the peak intensity was effectively restored to $3.08 \times 10^{21}\ \mathrm{W/cm^2}$ (Fig.~6(c)).

Consequently, this restoration led to a significant enhancement in the proton cutoff energy. Fig.~6(d) illustrates the dependence of proton energy on target defocus, where the zero position corresponds to the optimized focal plane at low-power. Before HotLoop correction, the peak proton energy occurred at a defocus of $-400\ \mathrm{\mu m}$, consistent with the previously observed longitudinal pre-focusing shift at full power. After HotLoop optimization, the maximum proton energy aligns with the zero-defocus position, confirming the successful compensation of the dominant defocus term. This recovery of focus quality leads to a 59\% enhancement in average proton energy (rising from 27\,MeV to 43\,MeV), demonstrating the significant impact of high-power wavefront correction on acceleration performance.

\section{Conclusion}
In conclusion, we have demonstrated the first successful implementation of the HotLoop strategy for optimizing 1 PW laser focal spots downstream of a plasma mirror. By utilizing a highly-attenuated, high-fidelity replica (the ``twin focus'') of the high-power focal spots, we quantified the increasing thermal aberrations with the rise of the laser power. The closed-loop correction of these distortions yielded a 2.67-fold enhancement in peak intensity. The efficacy of this approach was further validated by a 59\% enhancement in average proton cutoff energy during ion acceleration experiments.

Our work provides a robust solution for optimizing the performance of ultra-high intensity laser facilities through in-situ, closed-loop wavefront control. Beyond current applications, the HotLoop protocol is integrable with spatio-temporal diagnostics for single-shot 3D field reconstruction\cite{36,37,38}, offering a comprehensive characterization of the pulse structure. Additionally, it can serve as a unique testbed to characterize phase distortions in large-aperture transmissive optics (e.g., waveplates and debris shields) under full-power laser irradiation. Specifically, the large-aperture transmissive optics can be positioned in the collimated path of the DM chamber (see Fig.~1) using motorized stages. By differentially analyzing high-power wavefronts with and without the optic, both static distortions and dynamic thermal aberrations can be precisely characterized.

\section*{Acknowledgement}
This work was supported by the following projects: the National Science Fund for Distinguished Young Scholars (12225501), the National Key Research and Development Program of China (2024YFF0726304), the National Grand Instrument Project (2019YFF01014402), the Beijing Natural Science Foundation (1252019), the National Natural Science Foundation of China (12575257, 12595360,12595361,12595362), the Guangdong High Level Innovation Research Institute (2021B0909050006). Wavefront data analysis was performed on the High-Performance Computing Platform of Peking University.

\section*{Author Declarations}
\subsection*{Conflict of Interest}
The authors have no conflicts to disclose.

\section*{Data Availability}
The data that support the findings of this study are available from the corresponding author upon reasonable request.


\end{document}